\documentclass[manuscript,screen]{acmart}

\usepackage{natbib}
\usepackage{subfigure}
\usepackage{enumitem}
\usepackage{pifont}
\usepackage{tikz}
\usepackage{xcolor}
\usepackage[ruled,linesnumbered,vlined]{algorithm2e}
\makeatletter
\newcommand{\removelatexerror}{\let\@latex@error\@gobble}
\makeatother
\usepackage{booktabs}
\usepackage{adjustbox}
\usepackage{xspace}
\usepackage{subcaption}
\usepackage{enumitem}
\usepackage{pifont}
\usepackage{tikz}
\usepackage{xcolor}

\usepackage{array}
\usepackage{graphicx}
\usepackage{caption}
\usepackage{arydshln}
\usepackage{bbold}
\usepackage{mathtools}
\usepackage{multirow}

\renewcommand\footnotetextcopyrightpermission[1]{}

\usepackage[many,theorems,skins,breakable]{tcolorbox}
\newtcolorbox{visionbox}[2][]{%
    colback=teal!10,
    coltitle=black,
    colframe=teal!50,
    fonttitle=\bfseries,
    title=#2, 
    sharp corners,
    rounded corners=southeast,
    boxrule=0pt,
    enhanced,
    drop fuzzy shadow,
    #1, 
    top=3pt,bottom=2pt,left=3pt,right=3pt
    }

\usepackage[ruled,linesnumbered,vlined]{algorithm2e}

\newcommand{\algName}{\texttt{LEAD}\xspace}
\newcommand{\TORA}{\texttt{TORA}\xspace}
\newcommand{\LAF}{\texttt{LAF}\xspace}
\newcommand{\greedy}{\texttt{CD}\xspace}

\newcommand{\probName}{\texttt{RARS}\xspace}
\newcommand{\kit}{\texttt{E$^2$-RideKit}\xspace}

\definecolor{BBLUE}{rgb}{0, 0.4, 0.9}
\definecolor{MBLUE}{rgb}{0, 0.5, 0.8}
\definecolor{POrange}{rgb}{0.9, 0.5, 0.3}

\AtBeginDocument{%
  }

\setcopyright{acmlicensed}
\copyrightyear{2018}
\acmYear{2018}
\acmDOI{XXXXXXX.XXXXXXX}

\acmConference[FAccT’25]{ACM Conference on Fairness, Accountability, and Transparency}{June 23-26,
  2025}{Athens, Greece}
\acmISBN{978-1-4503-XXXX-X/2018/06}

\begin{document}

\title[LEAD: Equity-Aware Decarbonization of Ridesharing]{LEAD: Towards Learning-Based Equity-Aware Decarbonization\\ in Ridesharing Platforms}


\author{Mahsa Sahebdel}
\affiliation{%
  \institution{University of Massachusetts Amherst}
  \city{Amherst}
  \state{MA}
  \country{USA}
}
\email{msahebdelala@umass.edu}  

\author{Ali Zeynali}
\affiliation{%
  \institution{University of Massachusetts Amherst}
  \city{Amherst}
  \state{MA}
  \country{USA}
}
\email{azeynali@umass.edu}

\author{Noman Bashir}
\affiliation{%
  \institution{Massachusetts Institute of Technology}
  \city{Cambridge}
  \state{MA}
  \country{USA}
}
\email{nbashir@mit.edu}

\author{Prashant Shenoy}
\affiliation{%
  \institution{University of Massachusetts Amherst}
  \city{Amherst}
  \state{MA}
  \country{USA}
}
\email{shenoy@cs.umass.edu}

\author{Mohammad Hajiesmaili}
\affiliation{%
  \institution{University of Massachusetts Amherst}
  \city{Amherst}
  \state{MA}
  \country{USA}
}
\email{hajiesmaili@cs.umass.edu}


\begin{abstract}
Ridesharing platforms such as Uber, Lyft, and DiDi have grown in popularity due to their on-demand availability, ease of use, and commute cost reductions, among other benefits. 
However, not all ridesharing promises have panned out. 
Recent studies demonstrate that the expected drop in traffic congestion and reduction in greenhouse gas (GHG) emissions have not materialized. 
This is primarily due to the substantial distances traveled by the ridesharing vehicles without passengers between rides, known as deadhead miles. 
Recent work has focused on reducing the impact of deadhead miles while considering additional metrics such as rider waiting time, GHG emissions from deadhead miles, or driver earnings. 
However, most prior studies consider these environmental and equity-based metrics individually despite them being interrelated. 
In this paper, we propose a Learning-based Equity-Aware Decarabonization approach, \algName, for ridesharing platforms. 
\algName targets minimizing emissions while ensuring that the driver's utility, defined as the difference between the trip distance and the deadhead miles, is fairly distributed.
\algName uses reinforcement learning to match riders with drivers based on the expected future utility of drivers and the expected carbon emissions of the platform without increasing the rider waiting times. 
Extensive experiments based on a real-world ridesharing dataset show that \algName improves the defined notion of fairness by 150\% when compared to emission-aware ride-assignment and reduces emissions by 14.6\% while ensuring fairness within 28--52\% of the fairness-focused baseline. It also reduces the rider wait time, by at least 32.1\%, compared to a fairness-focused baseline.
\vspace{-0.2cm}
\end{abstract}





\definecolor{goldButNotOld}{HTML}{F26035}
\newcommand{\mo}[1]{\textcolor{blue}{[MH: #1]}}
\newcommand{\todo}[1]{{\textcolor{blue}{TODO: #1}}}
\newcommand{\mah}[1]{\textcolor{purple}{#1}}
\newcommand{\mahsa}[1]{\textcolor{black}{#1}}
\newcommand{\noman}[1]{\textcolor{teal}{\textbf{NB:} #1}}
\newcommand{\ali}[1]{\textcolor{goldButNotOld}{[Ali: #1]}}
\newcommand{\rev}[1]{\textcolor{red}{#1}}

\maketitle

\section{Introduction}
\label{sec:intro}
Road transportation significantly contributes to global energy consumption and greenhouse gas (GHG) emissions~\cite{davis2021transportation}. 
Worldwide, the transportation sector is the fourth-largest source of GHG emissions and is the largest in the United States~\cite{epa_transportation_ghg,sadati2021hybrid}. 
In 2021, transportation accounted for 37\% of global CO2 emissions~\cite{iea2022worldenergy}. 
That same year, the United States emitted approximately 5 billion metric tons of carbon dioxide, accounting for about 13.49\% of global emissions, exceeding the combined emissions of the 28 countries of the European Union~\cite{climategov2024co2emissions}. 
Due to these environmental impacts, there is an effort to explore transport systems that move people in an eco-friendly manner~\cite{schnieder2023effective}.

In recent years, shared mobility options have emerged as an attractive transport solution due to the widespread integration of smartphones into daily life. 
These services
enable users to access transportation on demand without owning a vehicle~\cite{shaheen2016shared, bongiovanni2022machine, ma2020spatio, ma2022general}. 
As a result, ridesharing is seen as a sustainable transportation option that can help reduce emissions and traffic congestion.
Unfortunately, these expectations have not materialized
and ridesharing services have been linked to increased traffic congestion, higher emissions, and negative social equity implications~\cite{Henao2017ImpactsOR, safaeian2023sustainable, barrios2023cost}. 
The increase in traffic congestion and emissions results from two key by-products of ridesharing. 
First, ridesharing availability can reduce the use of public transportation. 
Second, ridesharing vehicles travel a significant fraction of their total miles without a passenger on their way to pick up a passenger or return after dropping one~\cite{loa2023influences, shi2021influence, sahebdel2023data}. These non-passenger miles are referred to as \textit{deadhead} miles. Previous studies
show that deadhead miles can account for 19\% to 41\% of total miles while increasing emissions by up to 90\% compared to personal vehicle use~\cite{henao2019impact, wenzel2019travel}. 

Recent work has focused on optimizing deadhead miles in ridesharing ecosystems~\cite{kontou2020reducing, sahebdel2024holistic} while considering additional rider-specific metrics such as wait time~\cite{xu2019unified, lee2004taxi} or platform-wide objectives such as GHG emissions~\cite{sahebdel2024holistic, yan2020quantifying}. 
Other studies aim to improve overall rider satisfaction by minimizing travel costs and waiting times~\cite{cheng2017utility, levinger2020human, singh2021distributed}, but they often ignore fairness from the drivers' perspective, which can exacerbate social inequities. 
The studies from a driver's perspective typically focus on enhancing driver satisfaction and operational efficiency by optimizing routes and maximizing a given definition of utility~\cite{li2022ride, tang2019deep}. While some of those studies target fairness in earnings across drivers~\cite{shi2021learning}, they ignore the rider's waiting time and system-wide emissions. 
As a result, the ride-assignment algorithms proposed by prior work cannot be used in practice to promote sustainable and equitable growth in ridesharing platforms.

In this work, we consider the ride assignment problem that balances the objectives of the system (low emissions), the driver (fairness in earnings), and the rider (low waiting time). 
The goal is to minimize system-wide emissions, including emissions from ride and deadhead miles, while maximizing fairness in total utility across all trips between drivers without impacting the rider's waiting time. 
The utility of a trip for a driver is defined as the trip distance minus the deadhead distance. 
Given this setup,~\cite{sahebdel2024holistic} and~\cite{shi2021learning} represent our most relevant related works. 
In~\cite{sahebdel2024holistic}, the authors propose an online threshold-based ride assignment algorithm (\TORA) that targets minimizing the system-wide emissions while optimizing the waiting times. 
\TORA achieves a highly favorable tradeoff between emissions (up to 60\% reduction) and rider's wait time (4\% increase). 
However, to reduce emissions, \TORA assigns a disproportionate fraction of rides to low-emission vehicles, resulting in high unfairness among drivers. 
For example, it may assign up to 65\% of the rides to EVs despite them constituting only 25\% of the fleet. 
In~\cite{shi2021learning}, the authors propose \LAF that targets maximizing utility, defined as the total revenue generated by the system, while minimizing unfairness in the driver's equitable earnings. 
While taking a driver-centric approach, \LAF completely ignores system-wide emissions and the rider's wait time. 

To achieve the objectives of reducing system-wide emissions and ensuring fairness for drivers, we leverage reinforcement learning (RL).
RL's potential to handle sequential decisions and adapt based on feedback makes it a powerful tool for navigating dynamic and intricate environments. RL provides a framework in which agents learn to optimize their actions within an environment through trial and error, making it applicable to a wide range of application domains~\cite{sales2024efficient, han2022real, pigott2022gridlearn, wang2024incentive}.
In the setup of RL in the context of ridesharing, an agent performs an action that directly impacts the environment, which could include factors related to the city, drivers, or riders. In response, the agent receives feedback, such as the successful trip completion or the revenue earned, and uses this information to refine its future decision-making. Its ability to capture and understand the intricate spatial and temporal dynamics of demand patterns has shown great promise in solving operational inefficiencies commonly faced in ridesharing systems, such as vehicle dispatching, matching riders with drivers, and optimizing route planning in real time~\cite{cao2020using, asadi2021stochastic}. 

In this paper, we propose the \underline{L}earning-Based \underline{E}quity-\underline{A}ware \underline{D}ecarbonization (\algName) algorithm, which aims to minimize total system-wide emissions and maximize fairness in drivers' accumulated utilities across trips. 
To reduce system-wide emissions, \algName must consider the drivers available at present and the ones that will become available in the near future. 
To do so, \algName explicitly accounts for the dependencies among ride assignments by using reinforcement learning to develop future-aware ride assignment strategies. Using a real-world ridesharing dataset, we analyze how \algName and other representative baselines~\cite{sahebdel2024holistic, shi2021learning} perform using metrics, such as reduction in emissions, quality of service for riders measured as wait time, and fairness for drivers, and how gracefully they navigate the trade-offs between these metrics. 
Our contributions can be summarized as follows:

\begin{itemize}[leftmargin=*, itemsep=0.1cm]
    \item We frame the problem of ride assignment in ridesharing services (termed \probName) as a bipartite matching problem and propose a new objective function that aims to minimize total carbon emissions across all trips while reducing the gap between utilities of different drivers.
    
    \item We present \algName, an online learning-based algorithm that reduces emissions and ensures fairness for drivers, as measured by utility. To the best of our knowledge, this is the first study to leverage the dependencies between past and future ride assignments to minimize carbon emissions while ensuring a fair distribution of utility among drivers. 

    \item We implemented \algName on a simulation testbed and evaluated its performance using real data from RideAustin~\cite{rideaustin-dataset}, a nonprofit ridesharing service. In addition, we compare the performance of \algName with heuristic and state-of-the-art algorithms used in~\cite{sahebdel2024holistic, shi2021learning}. Our experimental results show that, in practical scenarios, our algorithm not only significantly improves the fairness of utility distribution among drivers but also substantially reduces emissions, compared to the current state-of-the-art methods.
\end{itemize}
\vspace{-0.3cm}

\section{Related Work}
\label{sec:related}
\noindent\textbf{Ridesharing Optimization Approaches.} The development of ridesharing systems has introduced unique challenges, with the primary focus being the assignment of ride requests to drivers. Some challenges in this area include improving vehicle utilization across multiple ride requests and demands \cite{xu2020efficient, luo2019dynamic,li2019top}, developing algorithms for route planning \cite{duan2020efficiently, wang2020demand}, and analyzing and integrating temporal and spatial patterns to predict the arrival time of riders' requests \cite{hong2020heteta, fang2021ssml}. Some other works in this domain also consider different optimization objectives such as maximizing drivers' profit \cite{wang2019adaptive, asghari2018adapt, zhang2017taxi}, while others aim to minimize travel costs \cite{tong2016online, wang2018stable} or reduce riders' waiting times \cite{yengejeh2021rebalancing, anthony2014online, xu2019unified}. 

In~\cite{kontou2020reducing} , authors attempt to reduce trip-level deadhead miles by leveraging hour-ahead trip demand predictions and using a heuristic approach to driver assignment. However, this focus on reducing trip-level deadhead miles does not always result in a reduction in system-wide deadhead miles and emissions, as these outcomes also depend on factors such as the fuel efficiency of vehicles and traffic conditions. Another work~ \cite{sahebdel2024holistic} focuses on minimizing emissions from deadhead miles while also reducing rider waiting times, ensuring no degradation in user satisfaction. However, a ride assignment system focused only on reducing emissions can negatively impact the quality of service for riders and drivers.
For example, a naive algorithm focused primarily on emission reduction may favor electric or low-emission vehicles over high-emission ones, leading to unfair ride assignments for drivers with high-emission vehicles. These drivers often belong to lower-income communities and may face diminished earning opportunities as a result.
Additionally, an emissions-aware ride assignment may allocate trips with longer deadhead miles to low-emission vehicles. This increases the deadhead-to-trip ratio, decreasing these vehicles' overall efficiency and service quality. 

Another key challenge is the lack of consideration for the temporal dependencies between current and future assignments. This oversight can impact the optimization of utility and fairness. 
 Many algorithms in this area attempt to provide theoretical performance guarantees, but they typically rely on myopic assumptions~\cite{xu2019unified}. For instance, they frequently neglect the dependency of assignments on past and future decisions, which can lead to sub-optimal assignments. This shortsightedness fails to account for the dynamic and evolving nature of ridesharing systems, where current ride assignments can significantly impact future states and the system's overall efficiency. A practical alternative approach involves leveraging historical data to optimize ride assignments alongside integrating reasonable assumptions. 

\vspace{0.05cm}
\mahsa{\noindent\textbf{Reinforcement Learning in Ridsharing.}  Numerous prior studies have highlighted the potential and advantages of utilizing reinforcement learning. Several comprehensive surveys on reinforcement learning (RL) for intelligent transportation systems exist, with most focusing on applications such as traffic signal control, autonomous driving, and routing optimization \cite{haydari2020deep, yau2017survey}. Song et al.~\cite{song2020application} conduct a case study on ridesharing in Seoul, utilizing a tabular Q-learning agent to optimize spatiotemporal pricing. Their extensive simulations focus on the issue of driver refusals to transport passengers to low-demand areas, which results in longer wait times for riders in these regions. Additionally, they investigate how surge pricing can mitigate the challenges faced by these marginalized zones—areas where taxis are typically scarce—and contribute to improving spatial equity throughout the city. Lin et al. \cite{lin2018efficient} developed a sophisticated contextual multi-agent reinforcement learning framework that combines contextual deep Q-learning with a contextual multi-agent actor-critic approach. This framework facilitates explicit coordination between a large number of vehicles operating on a large-scale, on-demand ridesharing platform. By incorporating both learning techniques, the framework is able to enhance the decision-making processes of each agent (vehicle) while enabling them to work together seamlessly. In another work, Teng et al. \cite{tang2019deep} developed a cerebellar value network for optimizing multi-driver order dispatching in ridesharing systems. This work aims to improve the stability of state value iteration when employing nonlinear function approximators and demonstrate that this new approach substantially enhances driver income and passenger experience metrics. Al-Abbasi et al. \cite{al2019deeppool} developed a model-free method leveraging deep Q-network (DQN) techniques to determine optimal dispatch policies through interaction with the environment in a ridesharing system. }

\vspace{0.05cm}
\noindent\textbf{Fairness in Ridesharing.} Fairness is a critical factor in matching scenarios, particularly in multi-sided markets where preferences play a significant role \cite{garfinkel1971improved, lorenz1905methods, dickerson2014price}. In the context of ridesharing, recent literature has raised concerns about the fairness of the assignments. Brown et al.~\cite{brown2018ridehail} shows the unfair treatment of riders by ridesharing companies, which leads to higher trip cancellation rates for a few groups of riders. Another study indicates that income inequality among ridesharing drivers may prevent them from earning a living wage~\cite{graham2017towards}. Another work aims to optimize long-term efficiency and fairness in ridesharing platforms through joint order dispatching and driver repositioning \cite{sun2024optimizing}. Another recent work~\cite{sahebdel2024holistic} showed a trade-off between reducing carbon emissions and sacrificing the fairness of utilities among drivers. Similarly, another study~\cite{shi2021learning} demonstrated optimization in driver utilities and fairness without considering emissions. This study used reinforcement learning techniques to evaluate the expected achievable utilities in different city areas. Despite the success of this approach in ensuring equity among drivers' utilities, it did not address eco-friendly ride assignments with low carbon emissions.

Ensuring fair assignments in ridesharing systems is challenging due to drivers' and riders' inherent spatiotemporal dynamics. Drivers can earn varying amounts of utility based on their assignments, while riders may experience differing waiting times. These disparities can lead to perceptions of unfairness, complicating the assignment process. To tackle these challenges, our work introduces a new fairness metric for driver earnings that considers the spatiotemporal dynamics inherent in ridesharing systems. This metric is designed to ensure a more equitable distribution of earnings among drivers, considering the dynamic nature of the system. Building on this metric, we have developed a reinforcement learning-based approach assignment algorithm that reduces carbon emissions while providing a fair distribution of driver utilities compared to the state-of-the-art assignment algorithm~\cite{sahebdel2024holistic}. The algorithm leverages historical data and dynamic learning to create effective and equitable assignment strategies, addressing immediate and long-term objectives in the ridesharing ecosystem. 

\section{Problem Formulation}
\label{sec:problem}
We consider a ridesharing system consisting of a set of drivers, denoted as $\mathcal{V}$, and a set of ride requests, denoted as $\mathcal{R}$. In the ridesharing context, a ride $r \in \mathcal{R}$ is defined as a request that includes a rider's pickup location, denoted as $p_r$, a rider's dropoff location, denoted as $q_r$, and time of the request, denotes as $c_r$. A driver $v \in \mathcal{V}$ in the ridesharing system is defined by their current location, utility, and the emissions per mile of the driver's vehicle $e_v$.

In ridesharing platforms, riders set the pickup and dropoff locations for a ride request; then, the platform matches the request to an available driver and estimates the trip duration and waiting time based on factors such as trip distance, traffic congestion, and the distance between the assigned driver and the rider's pickup location. 
An available driver in the system refers to a driver who is either idle or soon to be available within the specified time to serve a request. The emission associated with a ride request $r$ is influenced by the unit emissions of the driver's vehicle, the deadhead distance, the trip distance, and additional factors such as traffic congestion, road conditions, etc. For simplicity, we estimate the carbon emissions produced during the servicing of the ride request $r$ by driver $v$ using the equation below.
\begin{equation}
E_{v,r} = (d_{v,r}^{(D)} + d_{r}^{(T)}) \cdot e_v,
\end{equation}
where $e_v$ is the carbon emission per mile of the vehicle of driver $v$, $d_{v,r}^{(D)}$ is the deadhead distance of the trip, and $d_{r}^{(T)}$ is the trip distance which is the distance between rider's pickup and dropoff location. 
The utility of a driver $v$ is defined as the difference between the trip distance and deadhead distance accumulated across all served ride requests. 
The trip distance represents the actual distance driven to fulfill the ride request, which contributes to the driver's earnings. In contrast, the deadhead distance, the distance traveled without carrying a passenger (e.g., traveling from the driver's location to the rider's pickup location), incurs an overhead cost for the driver beyond the costs incurred during the trip, such as fuel expenses. 
By subtracting the deadhead distance from the trip distance, we effectively account for the revenue generated and the overhead cost incurred by the driver. This approach provides a measure of the driver's net revenue, reflecting the true utility of a ride to the driver:
\begin{equation}
U_{v,r} = \sum_{r' \in \mathcal{V}_r} ( d_{r'}^{(T)} - d_{v,r'}^{(D)}) \cdot x_{v,r'},
\end{equation}
where $U_{v,r}$ is the cumulative utility of the driver $v$ before serving ride request $r$, $\mathcal{V}_r$ is the set of ride requests posted before $r$, and $x_{v,r} \in \{0, 1\}$ is a decision variable: $x_{v,r} = 1$ if ride request $r$ is assigned to driver $v$; $0$ otherwise. For simplicity, we denote the utility of driver $v$ after serving all requests as $U_v$.

 



 


 
 

 



 
 

The ride-assignment problem builds on the classic bipartite matching problem, involving riders and drivers as distinct sets. The main challenge is modeling the constraints of available drivers and their costs for new requests, especially when some drivers are already occupied but may soon become available to take a ride. Our goal is to design an algorithm that assigns available vehicles to ride requests in a manner that minimizes emissions while ensuring fairness among drivers and considering the long-term effects and interdependencies of current assignments on future ones. We formulate the \probName problem as follows.
\begin{subequations}
\label{eq:LEAD_ILP}
    \begin{eqnarray}
    \label{eq:ILP_objective}
      [\probName] & \min &  \text{Emission} - \eta \cdot \text{Fairness}, \\
      && \text{Emission} := \sum_{v} \sum_{r} E_{v,r}\cdot x_{v,r},\\
      && \text{Fairness} :=-\max_v(U_{v}) + \min_v(U_{v}),\\
      \label{eq:constraint_a}
      &\textrm{s.t.,} & \sum_{v}  x_{v,r} \in \{0, 1\}  \qquad \forall r, \\
      \label{eq:constraint_b}
      && \sum_{\mathclap{\{r' \neq r | c_r \leq c_{r'} \leq  a_{r,v} \}}}  x_{v,r'} \cdot x_{v,r} = 0  \quad \ \forall v,r, \\
       \label{eq:constraint_c}
      &\textrm{vars.,} & \ \ \quad x_{v,r} \in \{0, 1\} \qquad  \forall v, r.
    \end{eqnarray}
\end{subequations}
where $a_{r,v}$ shows the dropoff time of ride request $r$ when served by driver $v$. The term \textit{Emission} includes trip emissions and deadhead miles emissions. The \textit{Fairness} metric is defined as the difference between the minimum and maximum utilities of the drivers and is a non-positive metric. Constraint~\eqref{eq:constraint_a} ensures that each ride request is assigned to at most one driver, and constraint~\eqref{eq:constraint_b} ensures that each driver is assigned to at most one ride request at each time. The parameter $\eta \geq 0$ controls the balance between emissions and fairness. Specifically, it quantifies the increase in emissions in grams of carbon dioxide (g.CO2) that we will accept for a one-kilometer decrease in unfairness across drivers. 
A higher value of $\eta$ indicates a greater emphasis on fairness. A lower value of $\eta$ prioritizes reducing emissions, potentially at the cost of sacrificing fairness.

\section{The \algName Algorithm}
\label{sec:method}
This section introduces the \algName algorithm that leverages reinforcement learning and batching to accumulate multiple ride requests that are jointly assigned to ensure that a long-term system-wide beneficial assignment is obtained.  
We present the formulation of the single batch matching problem as an integer linear programming problem whose objective includes two terms: one for the expected emission of ridesharing service and the second for the expected fairness in utility across drivers. 
To evaluate these expected values, \algName employs reinforcement learning to account for temporal dependencies between current and future ride-matching. This is essential since the system operates online, with rider requests and driver availability dynamically changing in real-time. 

In the following, we present the details of \algName, which consists of two main modules: the \textit{Learning Based} module and the \textit{Batched Emission Aware Fair Assignment} module. These modules are discussed in detail in Section~\ref{subsec:learn} and Section~\ref{subsec:alg}, respectively.
\vspace{-0.4cm}

\begin{figure}[t]
    \centering
    \includegraphics[width=0.7\linewidth, height=20cm, keepaspectratio]{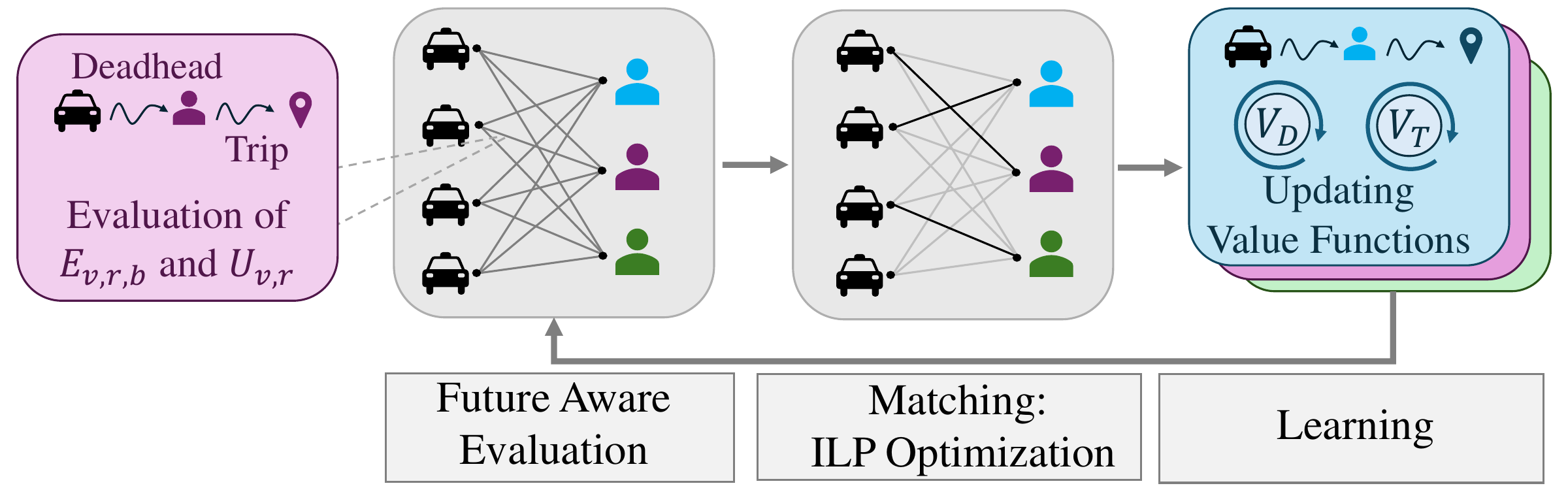}
    \vspace{-0.25cm}
    \caption{Overview of \algName. During each batch, \algName evaluates the long-term emissions and utilities and uses them to construct the weighting of the ILP problem of~\eqref{eq:LEAD_ILP}. After solving the optimization problem and matching ride requests with drivers, \algName updates the value functions based on the deadhead and trip distances of the served requests in the batch.}
    \vspace{-0.55cm}
    \label{fig:workflow} 
\end{figure}

\subsection{Learning Expected Emissions and Utility}
\label{subsec:learn}
In the following, we discuss the use of online reinforcement learning to model the impact of current assignments on future assignments and evaluate the expected deadhead and trip distances of trips in different regions of the covered area. 
Reinforcement learning is a method where agents learn by interacting with the environment over time. At each step, the agent takes an action and receives feedback in the form of utility from the environment. This continuous interaction allows the agent to optimize its strategy based on the utility received, making it well-suited for dynamic and complex systems such as ridesharing, where decisions made at one point can influence outcomes in the future. 

\mahsa{In this work, we focus on learning value functions to enable future-aware ride assignment. Common model-free algorithms, such as Temporal Difference (TD) learning~\cite{sutton1988learning}, Q-learning~\cite{watkins1992q}, and Actor-Critic methods~\cite{konda1999actor}, can be leveraged for this task, each with its strengths and limitations. Q-learning, which estimates action-value functions, faces challenges in our case due to insufficient data for each state-action pair. Actor-Critic methods, which learn both a value function and a policy simultaneously (with the Critic estimating the value and the Actor updating the policy based on feedback), are also unsuitable as the policy is dictated by riders and cannot be altered. This coupling of policy and value functions limits their use in our context. Given these constraints, we adopt TD learning for our work. TD learning updates value estimates incrementally, making it well-suited to our data limitations. The key elements of our reinforcement learning module are as follows:} 

\begin{itemize}[leftmargin=*]
    \item \textbf{Agent: }Each active driver $v$ in the ridesharing system is considered as an agent.
    \item \textbf{State:} The state of driver $v$ during batch $b$ is represented by the driver's location $l_{v,b}$. To facilitate data analysis and management, city space is discretized into a square grid system, which results in a finite number of unique states.
    \item \textbf{Action:} The action of the available driver is defined as the request they choose to accept.
\end{itemize}
After taking action, each agent receives utility, which is the difference between the trip distance and the deadhead distance. To predict the expected emissions produced and the utility achieved by a driver, we introduce two value functions for the agent: one for the deadhead distance and another for the trip distance.

Due to the complexity of considering multiple dependent agents within the system,
we simplify the learning module by treating different agents as a single agent. In the emission-aware fair assignment module, we leverage the evaluated value functions to take appropriate actions for the different action-correlated agents. We adopt a framework that involves sequences comprising the current state, current action, reward, and next state to estimate the value functions. \mahsa{For learning in this scenario, we use TD learning which iteratively updates value functions based on the deadhead and trip distances of each completed ride request.}
Below, $V_D(s)$ shows the expected deadhead distance a driver could travel when starting from state $s$.
\begin{equation}
\label{eq:valuef_deadhead}
V_D(s) = \mathbb{E}\left[\sum_{t=0}^{\infty} \gamma^t \cdot d_{t, D} | s_0 = s\right],
\end{equation}
where $d_{t, D}$ is the deadhead distance of $t^{th}$ trip served by the agent, $s_0$ denotes the starting state of the agent, and $\gamma \leq 1$ is the discount factor. Similarly, we define $V_T(s)$ as the expected trip distance a driver will travel starting from state $s$.
\begin{equation}
\label{eq:valuef_trip}
V_T(s) = \mathbb{E}\left[\sum_{t=0}^{\infty}  \gamma^t \cdot d_{t, T} | s_0 = s \right],
\end{equation}
where $d_{t, T}$ is the trip distance of $t^{th}$ trip served by the agent. Now, assume the agent is located in state $s_t$ and after serving the request, travels to the next state $s_{t+1}$. The TD learning update rule for the value functions based on the trip and deadhead distance of the trip are as follows:
\begin{align}
\label{eq:update_VD}
V_D(s) \leftarrow& V_D(s) + \alpha \left[d_{t, D} + \gamma V_D(s_{t+1}) - V_D(s_t)\right],\\
\label{eq:update_VT}
V_T(s) \leftarrow& V_T(s) + \alpha \left[d_{t, T} + \gamma V_T(s_{t+1}) - V_T(s_t)\right],
\end{align}
where $\alpha$ is the learning rate. The values of $V_D(s)$ and $V_T(s)$ are initialized through a specific initialization process (such as zero initialization or random initialization). After a driver serves a ride request, the values of these functions are updated accordingly. In the next section, we demonstrate how these value functions can be used to evaluate the expected deadhead and trip distances for trips starting in different city locations, ultimately estimating the future emissions and utility of drivers.
\vspace{-0.3cm}
\subsection{Batched Emission-Aware Fair Assignment}
\label{subsec:alg}
This section proposes matching ride requests to available drivers within \algName. \algName performs the ride matching process in a batch setting, where each batch comprises ride requests posted after the assignments from the previous batch. Let \(E_{v, r, b}\) denote the long-term expected emission produced by driver \(v\) after being assigned to request \(r\) within batch \(b\). The value of \(E_{v, r, b}\) depends on the trip and deadhead distances for serving request \(r\), as well as future deadhead and trip distances traveled by driver \(v\). To estimate the latter, we leverage the value function described in the previous section. Specifically, the expected emission produced by driver \(v\) during and after batch \(b\) can be calculated as: 
\begin{align}
E_{v,r,b} &= \mathbb{E}\left[\sum_{t=0}^{\infty} \gamma^t (d_{t, T} + d_{t, D}) \cdot e_v \big| s_0 = l_{v, b}, s_1 = q_r \right] \notag\\
\label{eq:weight_expected_emission}
&= \bigg[ (d_{r}^{(T)} + d_{v,r}^{(D)}) + \gamma (V_T(q_r) + V_D(q_r)) - (V_T(l_{v,b}) + V_D(l_{v,b})) \bigg] \cdot e_v,
\end{align}
which is directly derived from Equations~\eqref{eq:valuef_deadhead} and \eqref{eq:valuef_trip}, considering that the driver is located at state $l_{v,b}$ and will travel to the dropoff location of the request after serving of that. Similarly, let \(U_{v,r,b}\) denote the expected final utility of driver \(v\) given that they were assigned to request \(r\) during batch \(b\) and had a utility history of \(U_{v,r}\) by that time. We can evaluate the value of \(U_{v,r,b}\) as follows:
\begin{align}
U_{v,r,b} &= \mathbb{E}\left[\sum_{t=0}^{\infty} \gamma^t (d_{t, T} - d_{t, D}) \big| s_0 = l_{v, b}, s_1 = q_r \right] \notag\\
\label{eq:expected_utility}
&= U_{v,r} + (d_{r}^{(T)} - d_{v,r}^{(D)}) + \gamma (V_T(q_r) - V_D(q_r)) - (V_T(l_{v,b}) + V_D(l_{v,b})).
\end{align}

During the ride-matching of batch $b$, \algName finds the set of requests $\mathcal{R}_b$, and available drivers in that batch, $\mathcal{V}_b$, and evaluates $E_{v,r,b}$, and $U_{v,r,b}$ for every pair of $v$ and $r$ in the batch. Then, \algName finds the solution of the integer linear programming problem below to match each ride request to available drivers.
\begin{subequations}
\label{eq:LEAD_ILP}
    \begin{align}
    \label{eq:ILP_objective}
      \min \quad& \text{Emission}_{(b)} - \eta \cdot \text{Fairness}_{(b)}, \\
      & \text{Emission}_{(b)} := \sum_{v\in \mathcal{V}_b} \sum_{r \in \mathcal{R}_b} E_{v,r,b}\cdot x_{v,r},\\
          & \text{Fairness}_{(b)} := -\max_v \sum_{r \in \mathcal{R}_b}(U_{v,r,b} \cdot x_{v,r}) + \min_v\sum_{r\in \mathcal{R}_b}(U_{v,r,b} \cdot x_{v,r}),\\
      \label{eq:problem_1bitrate}
      \textrm{s.t.,}\quad& \sum_{v \in \mathcal{V}_b}  x_{v,r} \in \{0, 1\} \quad \forall r, \\
      &  \sum_{r\in \mathcal{R}_b} x_{v,r} \in \{0, 1\}  \quad \forall v, \\
       \label{eq:problem_decision_range}
      \textrm{vars.,} \quad&  \quad \quad \  x_{v,r} \in \{0, 1\} \quad \forall v, r,
    \end{align}
\end{subequations}
where $\text{Emission}_{(b)}$ represents the expected total emissions produced by the ridesharing service during and after batch $b$. $\text{Fairness}_{(b)}$ estimates the expected fairness of the system given the assignments in batch $b$. It is worth mentioning that if the number of ride requests in batch $b$ exceeds the number of available drivers, only the first $|\mathcal{V}_b|$ ride requests will be assigned, with the remaining requests being moved to the next batch.


The pseudocode for \algName is outlined in Algorithm~\ref{alg:alg_pseu}. Initially, \algName evaluates the values of $E_{v,r,b}$ and $U_{v,r,b}$ for every pair of $v$ and $r$ in batch $b$ (Lines 1-3). Next, it finds the solution for the optimization problem in \eqref{eq:LEAD_ILP}, determining which driver should serve each ride request. The final step is to update the value functions based on the actual trip and deadhead distances of the served ride requests within this batch. Specifically, for each request $r$ in the batch, \algName updates the value functions associated with trip and deadhead distances using the update rules presented in Equations~\eqref{eq:update_VD} and \eqref{eq:update_VT} (Lines 5-10). By integrating emission and fairness into a single objective, \algName balances reducing carbon emissions with the fairness of utilities among drivers. Additionally, by considering the long-term expected emissions and utilities instead of just the current values, \algName optimizes these metrics with a forward-looking perspective. 
The time complexity of LEAD depends on the size of the optimization problem per batch. While the problem size may grow with the number of drivers, practical techniques can effectively manage this complexity. For example, the optimization can be limited to drivers within a fixed range of the rider, as online assignment algorithms generally exclude distant drivers due to the high deadhead miles involved. This approach significantly reduces the problem size while maintaining solution quality.
\vspace{-0.5cm}

\begin{figure}[!t]
      \removelatexerror
      \begin{algorithm}[H]
      \label{alg:alg_pseu}
\SetCustomAlgoRuledWidth{0.45\textwidth} 
        \caption{\small LEAD($\mathcal{R}_b$, $\mathcal{V}_b$)}
\ForAll{$v \in \mathcal{V}_b, r \in \mathcal{R}_b$}
{
        Calculate $E_{v,r,b}$ using Equation~\eqref{eq:weight_expected_emission}\;
        Calculate $U_{v,r,b}$ using Equation~\eqref{eq:expected_utility}\;
      }
Matched-Pairs $\leftarrow$ Solution of optimization problem~\eqref{eq:LEAD_ILP}\;
\ForEach{$r \in \mathcal{R}_b$}
    { $v$ $\leftarrow$ Matched-Pairs($r$)\; 
    $s_t \leftarrow$ current location of $v$\;
    $s_{t+1} \leftarrow$ dropoff location of the request $r$\;
    Update $V_D(s_t)$ using Equation~\eqref{eq:update_VD}\;
    Update $V_T(s_t)$ using Equation~\eqref{eq:update_VT}\;
    }
    \textbf{return} Matched-Pairs\;
    \end{algorithm}
\end{figure}
\setlength{\textfloatsep}{0pt}
\setlength{\intextsep}{0pt}


\section{Experimental Evaluation}
\label{sec:evaluation}
In this section, we conduct comprehensive experiments to evaluate the performance of \algName using several metrics, including emissions per trip, fairness in driver's utility, and average wait times for riders. 
We compare \algName against state-of-the-art algorithms across a range of values for key configuration paremeters of batch durations, the fairness parameter $\eta$, and the percentages of low-emission vehicles. 
The following outlines the key questions the evaluation addresses and summarizes our findings.

\begin{enumerate}[leftmargin=*, topsep=0.3cm, itemsep=0.2cm]
	\item[\textbf{Q1}] How much emissions does \algName reduce, especially compared to other baselines targeting emissions reduction?
     \\ \textbf{Outcome:} \emph{\algName reduces emissions by at least 56.6\%, 23.5\%, and 8.4\%, compared to baseline algorithms that target improving fairness (\LAF~\cite{shi2021learning}), reducing deadhead miles (\greedy), and minimizing carbon emissions (\TORA~ \cite{sahebdel2024holistic}), respectively.
     \footnote{
     \autoref{sec:experimental_setup} describes the baseline algorithms in detail.} This substantial decrease highlights \algName's effectiveness in minimizing environmental impact (see~\autoref{subsec:per_trip_emissions}).}
    

    \item[\textbf{Q2}] How much improvement in fairness does \algName provide compared to the baseline targeting emissions reduction?
    \\ \textbf{Outcome:} \emph{\algName demonstrates a 150\% improvement in fairness compared to \TORA that balances minimizing system-wide emissions against the passenger wait times but does not explicitly target fairness across drivers  (see~\autoref{subsec:drivers_fairness}).}
    
    \item[\textbf{Q3}] How does \algName's rider wait time performance compare to the baseline algorithm that explicitly considers fairness? 
    \\ \textbf{Outcome:} \emph{\algName achieves 39.4\% improvement over \LAF while also ensuring that not degrading fairness for drivers by more than 28.0\% (see Sec.~\ref{subsec:riders_wait_time}).} 
\end{enumerate}
\vspace{-0.4cm}



\subsection{Experimental Setup}
\label{sec:experimental_setup}



\noindent \textbf{Parameter settings.} 
\mahsa{We evaluate the performance of \algName and baseline algorithms across a range of experimental settings. Specifically, we evaluate the impact of parameter $\eta$ by varying its values between 0.1~gCO2/km and 10~gCO2/km. 
We also evaluate the impact of different batch durations by adjusting the batch duration values between 2 and 10 minutes.  
To define unique states, we divide the city into square tiles, each with a width of 1~km. The value functions for the reinforcement learning model are initialized to 0.  We set the discount factor, $\gamma =0.9$, and the learning rate $\alpha = 0.025$ in line with the evaluation settings used in previous research~\cite{shi2021learning}, ensuring consistency in the comparative analysis. To introduce additional metrics for evaluating the system-level performance of \algName compared to baseline algorithms, we classify vehicles emitting less than 135 gCO2/km as low-emission (LEVs) and those emitting more than 270 gCO2/km as high-emission (HEVs).} 

\noindent
\textbf{Ridesharing dataset.}
\label{sec:exp_dataset}
Our experiments use a publicly available dataset from RideAustin~\cite{rideaustin-dataset}, a non-profit ridesharing service based in Austin, Texas. This dataset includes approximately 1.5 million trips over ten months in 2016 and 2017, collected
in Austin. It contains detailed trip information such as pick-up and drop-off coordinates of each trip, vehicle make and model, and distances traveled before, during, and after each trip.


\mahsa{In our experiments, we use a subset of the RideAustin dataset, focusing on trips from December 1, 2016, to December 15, 2016. This subset includes 58,866 ridesharing trips served by 150 drivers. We augment this dataset by including carbon emission data for both deadhead miles and individual trips. Additionally, we include equity information for both drivers and riders, using \kit~\cite{sahebdel2024holistic}, a publicly available toolkit designed for holistic evaluation of emissions in ridesharing platforms. We assume that riders have an expectation regarding the maximum time they are willing to wait before being assigned to a driver. Furthermore, drivers available within a 15-minute window are included in the set of available drivers for each batch.  
}
To assess the impact of replacing HEVs with LEVs, we created three different variants of original dataset. In each variant we randomly converted a certain percentage of the non-LEVs into electric vehicles (EVs). This process resulted in three datasets  where the proportion of LEVs was increased to 10\%, 15\%, and 25\%, respectively. We set the EV's emission intensity to 63.35 gCO2 per kilometer.\footnote{We used the Tesla Model Y as a representative EV, which has an energy efficiency rating of 16.9 kWh/100km. The unit distance emissions were calculated using the average carbon intensity of 408 gCO2eq/kWh for Austin, Texas.}

\vspace{0.05cm}
\noindent
\textbf{Comparison algorithms.}
\label{sec:comp_algs}
\mahsa{We evaluate the performance of \algName by comparing it against a heuristic baseline algorithm as well as two state-of-the-art algorithms from prior works, which are outlined in detail below.}
\begin{itemize}[leftmargin=*]
    \item \textbf{Closest Driver (CD)}: 
    \mahsa{Within a given batch of requests, this algorithm sequentially matches ride requests in the batch  to the nearest available driver. The matching process is conducted in a greedy manner, aiming to minimize the deadhead miles and reduce the rider's wait time. Although this algorithm does not explicitly account for emissions during the assignment process, the focus on minimizing deadhead distances indirectly helps control ride emissions. }
    \item \textbf{\LAF}~\cite{shi2021learning}: 
    This algorithm is designed to ensure a fair distribution of utility among drivers. The ride assignment process in \LAF consists of four key stages:  evaluating, assigning, guiding, and learning. 
    Initially, the algorithm begins with evaluating a bipartite graph where edge weights represent trip prices between passengers and drivers. These weights are updated to reflect both current and future earnings based on historical trip prices observed in the area near the pick-up and the drop-off locations. 
    During the assigning stage, \LAF utilizes a bi-objective assignment algorithm. This algorithm aims to balance the aggregate utilities of drivers and the equity in their utilities. 
    The learning stage involves refining weight updates through reinforcement learning. Finally, during the guiding stage, idle drivers will be guided to high-demand areas for their future assignments.
    
    Since \LAF aims to maximize total utilities and ensure equity in utilities, it uses a single value function to track the expected achievable utilities in different city regions. For adaptation to the current work, we set the utility (or price) of a ride request $r$ served by driver $v$ as the difference between the trip and deadhead distances and evaluated weights in Algorithm~1 of~\cite{shi2021learning} accordingly. Additionally, during our experiments, the weight of trip utilities over different times of the day is the same.

    \item \textbf{\TORA}~\cite{sahebdel2024holistic}: \TORA is an online ride-assignment algorithm that aims to balance passenger waiting times with system-wide emissions. To the best of our knowledge, it is the only existing online ride assignment algorithm that explicitly considers emissions. \TORA aims to reduce passenger waiting times by initially selecting the closest available driver. It then evaluates alternative available drivers by comparing their distances and the additional deadhead emissions they would generate relative to the closest driver. For each driver, \TORA calculates an  ``Emission-to-Distance'' (E2D) ratio, which is the ratio of the difference in deadhead emissions between a given driver and the closest driver to the difference in distances to the passenger between the two drivers. \TORA selects the most suitable driver based on these E2D values. Despite the consideration of carbon emissions during the process of ride matching, \TORA does not consider fairness in driver's utilities.
    
\end{itemize}

\noindent
\textbf{Performance metrics.}
We evaluate the performance of \algName and comparison algorithms using three primary performance metrics: \textit{per trip emissions}, \textit{fairness in driver's utility}, and \textit{rider waiting time}. These metrics provide a multifaceted view of how each algorithm performs across critical aspects of ridesharing services—environmental impact, driver equity, and the quality of experience of the riders. For fairness, we normalize our results to \LAF algorithm's outcomes when comparing different algorithms. However, in analyzing the effect of LEV penetration, we normalize the values to outcomes achieved by the LEAD algorithm without the addition of LEVs. 

\begin{figure*}[t]
\centering
\begin{tabular}{ccc}
\includegraphics[width=0.321\linewidth]{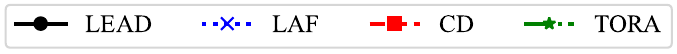}& \multicolumn{2}{c}{\includegraphics[width=0.55\linewidth]{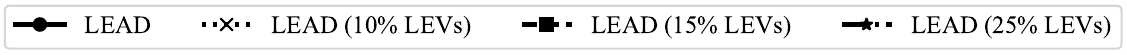}}\\
\includegraphics[width=0.3\linewidth]{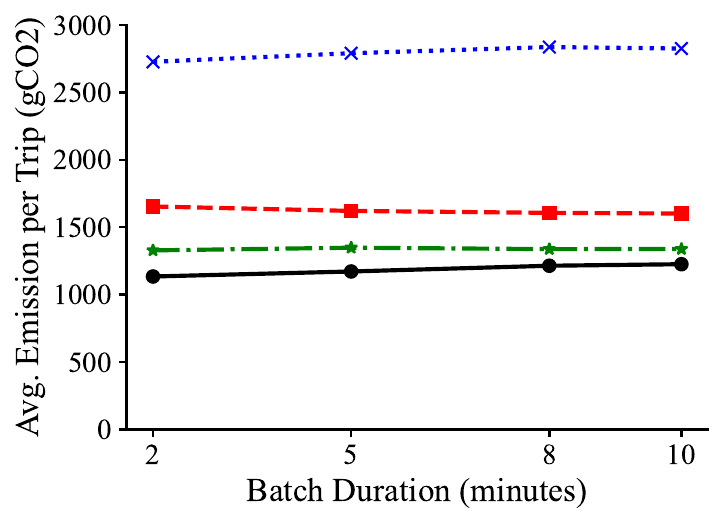} &  \includegraphics[width=0.3\linewidth]{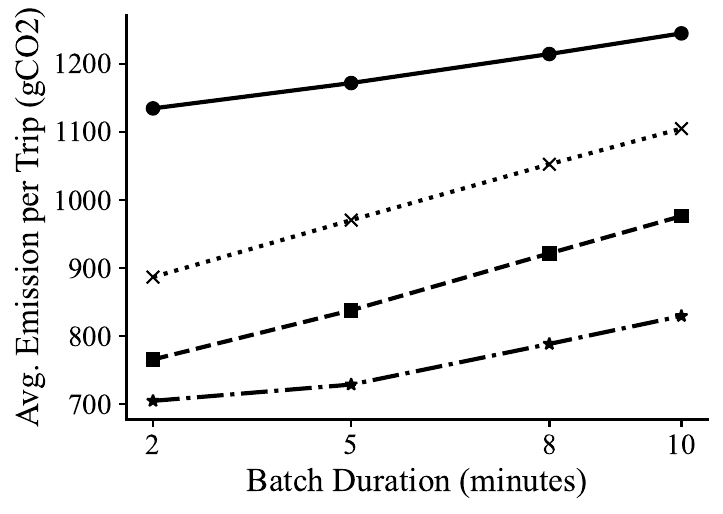} & \includegraphics[width=0.3\linewidth]{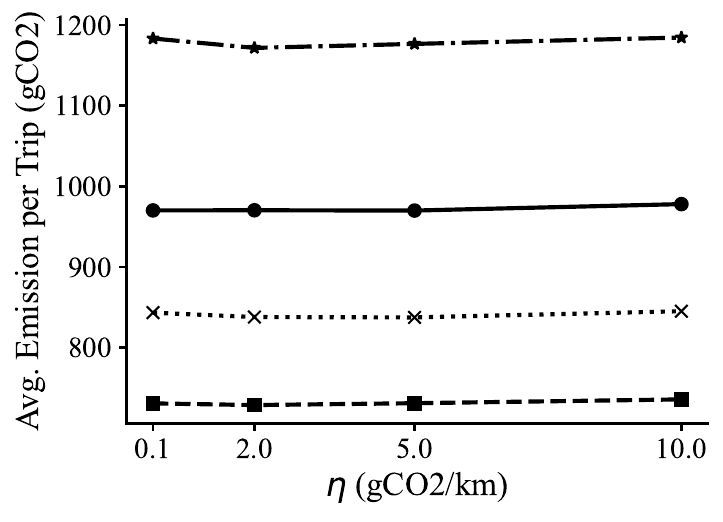}\\
  (a) \algName vs. baselines                &    (b) Impact of LEVs      &  (c) Impact of parameter $\eta$       
\end{tabular}
    \vspace{-0.3cm}
    \caption{Emissions reduction performance: (a) emission per trip for different algorithms as a function of batch duration, (b) impact of the percentage of low emission vehicles (LEVs) in the fleet on \algName for different batch durations and $\eta$ = 5, and (c) the impact of increasing emissions for an increase in fairness, captured using $\eta$, for a batch duration of 5 minutes. Here, $\eta$ specifies extra emissions that the algorithm incurs to reduce unfairness by 1km.
    }
    \label{fig:emissions} 
\end{figure*}


\vspace{-0.2cm}
\subsection{Impact on Per-Trip Emissions}
\label{subsec:per_trip_emissions}

\vspace{0.05cm}
\noindent
\textbf{Key results.}
We first outline the key observations here and defer the insights and analysis of findings to the next part. 
\mahsa{As shown in~\autoref{fig:emissions}~(a), \algName outperforms all the baseline algorithms across all different batch durations. The effectiveness of \algName in reducing emissions per trip is evident, with reductions of at least 23.5\% compared to the \greedy algorithm, 8.4\% compared to the \TORA algorithm, and 56.6\% compared to the \LAF algorithm.
Notably, \algName's performance worsens as the batch duration increases. 
While the emissions per trip for  \greedy and \TORA slightly decrease with an increase in batch duration, \algName achieves higher reductions at shorter batch duration, thereby increasing the margin of improvement. 
The maximum improvement of \algName compared to \TORA, \greedy, and \LAF occurs at a batch duration of 2 minutes, with improvements of 14.6\%, 31.4\%, and 58.4\%, respectively.
Finally, \LAF's performance worsens with an increase in batch duration, e.g., emissions are 3\% and 1\% higher at 10 minutes batch duration than at 2 and 5 minutes batch duration.}

As shown in~\autoref{fig:emissions}~(b), adding low-emission vehicles (LEVs) to the ridesharing fleet leads to a reduction in the average emissions per trip.  
Specifically, adding 10\% LEVs results in a reduction of emissions by up to 21.8\%.  Increasing the proportion of LEVs to 15\% reduces emissions by up to 32.5\%. When the proportion of LEVs is increased to 25\%, the emissions reduction achieves up to 37.9\% demonstrating a significant improvement in environmental performance.

\autoref{fig:emissions}~(c) shows that changing the fairness parameter $\eta$ has only a minor impact on the emissions.  The maximum increase in emissions per trip is observed to be 0.8\%, and this occurs when 10\% of the vehicles are LEVs. As we increase $\eta$ to 10gCO2/km, the emissions only increase by 0.21\% and 0.67\% at 15\% and 20\% LEV percentages, respectively. 

\vspace{0.1cm}
\noindent
\textbf{Analysis of findings.} We next delve deeper into some results observed about \algName's impact on emissions.
\begin{itemize}[leftmargin=*, topsep=-0.02cm]
    \item \textbf{Why does \algName outperforms deadhead mile- and emission-aware algorithms \greedy and \TORA?} 
    
    The primary reason that \algName outperforms \greedy is that CD does not take into account the unit emissions of the vehicles when assigning requests to drivers. This omission means that CD prioritizes minimizing the distance between drivers and riders without factoring in the environmental impact of each vehicle's emissions. When there is a wide variation in the emission rates across vehicles, the performance gap between \greedy and emission-aware algorithms like \algName and \TORA becomes more pronounced. Another reason that LEAD outperforms \TORA and \greedy is its approach for the assignment of multiple requests within a batch. It assigns all the requests in each batch simultaneously, whereas \TORA and \greedy assign the requests sequentially within each batch. This sequential approach limits their ability to optimize for emissions across the entire batch, as decisions are made one at a time rather than considering the batch as a whole.
    

    \item \textbf{Why does increasing the batch duration from 2 to 10 minutes increase emissions but improve fairness?}
    
    The RideAustin dataset includes many rides with short trip distances. When the batch duration is short (e.g., 2 minutes), the system assigns requests more frequently allowing it to repeatedly assign rides to low-emission vehicles. This frequent assignment of low-emission vehicles reduces emissions but leads to unfairness since high-emission vehicles are underutilized. When the batch duration increases (e.g., to 10 minutes), the system processes more requests at once, however assigns only a subset of these to low-emission vehicles. The remaining requests are distributed among high-emission vehicles. This results in a higher average emission but fairer allocation of rides between low-emission and high-emission vehicles.
    
    \item \textbf{Why don't the emissions per trip for \algName increase as $\eta$ rises from $0.1$ to $10$ gCO2/km?} 
    
    The value of $\eta$ shows the grams of CO2 emitted for each kilometer increase in fairness. Within that range of values for $\eta$, \algName carefully balances this tradeoff and only chooses the trips when a meaningful improvement in fairness can be achieved.
    
    
\end{itemize}

\vspace{-0.1cm}
\begin{visionbox}{}
\noindent
\textbf{Key takeaways.} 
\emph{\algName achieves up to 14.6\% emissions reduction compared to the state-of-the-art emission-aware algorithm \TORA. While additional reductions may be possible by optimizing parameters such as batch duration and $\eta$, it will introduce tradeoffs with our other objectives of increasing driver's fairness and reducing rider's wait time.}
\end{visionbox}
\vspace{-0.1cm}

\subsection{Impact on Driver's Fairness}
\label{subsec:drivers_fairness}

\begin{figure*}[t]
\centering
\begin{tabular}{ccc}
\includegraphics[width=0.321\linewidth]{Figures/algo_labels.pdf}& \multicolumn{2}{c}{\includegraphics[width=0.55\linewidth]{Figures/ev_labels.pdf}}\\
\includegraphics[width=0.3\linewidth]{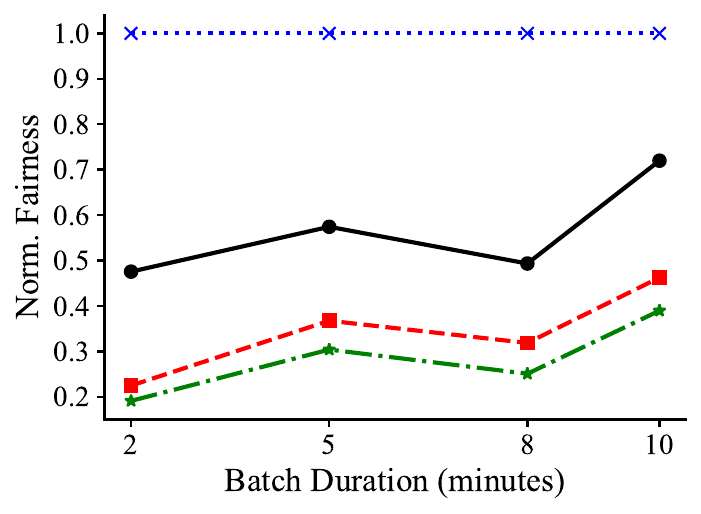} &  \includegraphics[width=0.31\linewidth]{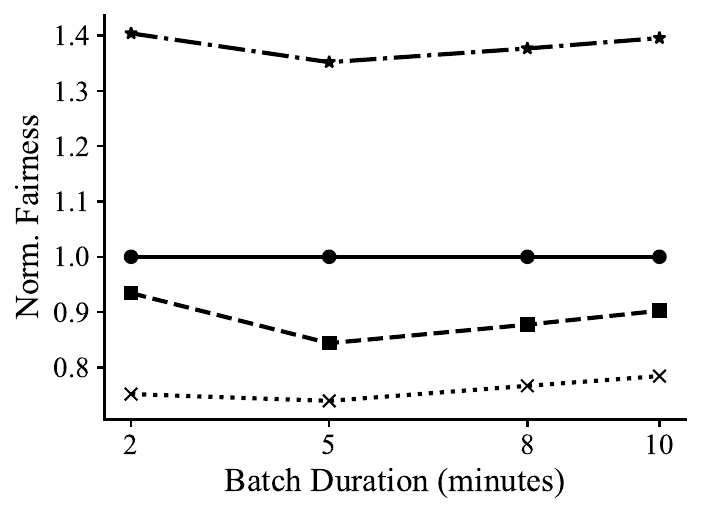} & \includegraphics[width=0.31\linewidth]{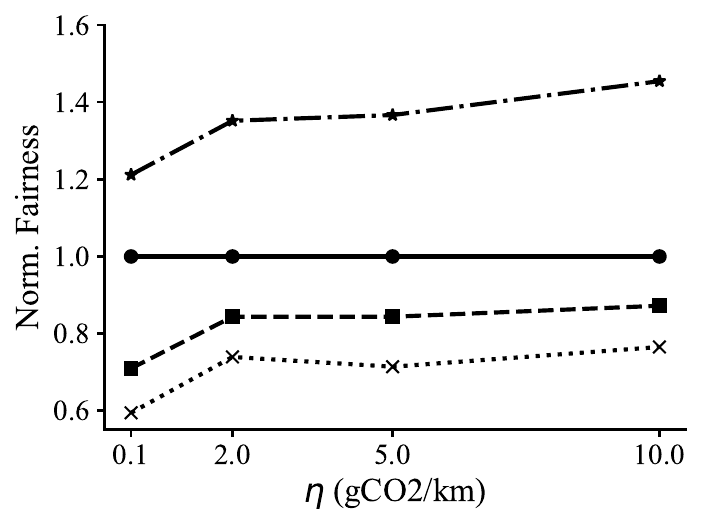}\vspace{-0.1cm}\\
  (a) \algName vs. baselines                &    (b) Impact of LEVs      &  (c) Impact of parameter $\eta$       
\end{tabular}
    \vspace{-0.2cm}
    \caption{Fairness performance: (a) normalized fairness for different algorithms as a function of batch duration, (b) impact of the percentage of low emission vehicles (LEVs) in the fleet on \algName for different batch durations and $\eta$ = 5, and (c) the impact of increasing emissions for an increase in fairness, captured using $\eta$, for a batch duration of 5 minutes. Here, $\eta$ specifies extra emissions that the algorithm incurs to reduce unfairness by 1km.
    }
    \label{fig:fairness} 
\end{figure*}

\noindent
\textbf{Key results.} 
As shown in~\autoref{fig:fairness}~(a), no other algorithm, including \algName, beats \LAF in terms of fairness performance. This happens since \LAF only prioretize fairness in its assignments. The next best fairness performance is shown by \algName, as it reaches 71.9\% of the fairness achieved by \LAF at the batch duration of 10 minutes. Meanwhile, \TORA is the worst performing algorithm as it achieves between 19.1\% to 39.0\% of \LAF's fairness performance at batch durations of 2 minutes and 10 minutes, respectively.

As shown in~\autoref{fig:fairness}~(b), incorporating 25\% LEVs has the most significant impact on fairness. Specifically, with 25\% LEVs, fairness improves by 40\% for batch durations of 2 minutes compared to the baseline ridesharing fleet with \algName. In contrast, increasing the proportion of LEV to 10\% or 15\% results in a decrease in fairness over all batch durations.

\autoref{fig:fairness}~(c) shows the effect of $\eta$ on fairness, providing us with two key observations. First, increasing the value of $\eta$ increases fairness. For example, at an $\eta$ value of 0.1 gCO2/km when 25\% of the fleet consists of LEVs, fairness increases by 21\%, while a much higher $\eta$ value of 10 gCO2/km leads to an 45\% increase in fairness. Second, fairness decreases at lower LEV percentages but improves significantly at higher percentages. For example, fairness improved by at least 60\% and up to 70\% when the percentage of LEVs was increased from 15\% to 25\%.


\noindent
\textbf{Analysis of findings.} We next look deeper at the results we presented to quantify the fairness properties of \algName.
\begin{itemize}[leftmargin=*]

   \item \textbf{What causes the significant increase in fairness when the proportion of LEVs rises from 15\% to 25\% compared to baseline \algName?} 
    
    Fairness increases with the addition of 25\% LEVs compared to the baseline \algName due to a availability of more number of LEVs. When the proportion of LEVs is 10\% or 15\%, the system prioritizes assigning most ride requests to LEVs to reduce emissions, which leads to unfairness. However, with 25\% LEVs, the system achieves a better balance in ride assignments between LEVs and HEVs, resulting in significantly improved fairness.
   

    \item \textbf{How increasing $\eta$ to improve fairness impacts emissions?} 
    
    As outlined in the previous section, changing $\eta$ between $0.1$ to $10$ gCO2/km has only a marginal impact on emissions, with a maximum increase of 0.8\%. Since increasing $\eta$ can lead to significant improvements in fairness by almost 70\%, the tradeoff seems desirable.  
    
    
\end{itemize}

\begin{visionbox}{}
\noindent
\textbf{Key takeaways.} 
\emph{\algName offers a highly favorable tradeoff between improving fairness and reducing emissions. For example, when compared to \LAF, at batch duration of 10 minutes, it reduces emissions by 56.6\% while degrading fairness by~29\%.}


\end{visionbox}

\subsection{Impact on Rider's Wait Time}
\label{subsec:riders_wait_time}
\begin{figure*}[t]
\centering
\begin{tabular}{ccc}
\includegraphics[width=0.321\linewidth]{Figures/algo_labels.pdf}& \multicolumn{2}{c}{\includegraphics[width=0.55\linewidth]{Figures/ev_labels.pdf}}\\
\includegraphics[width=0.3\linewidth]{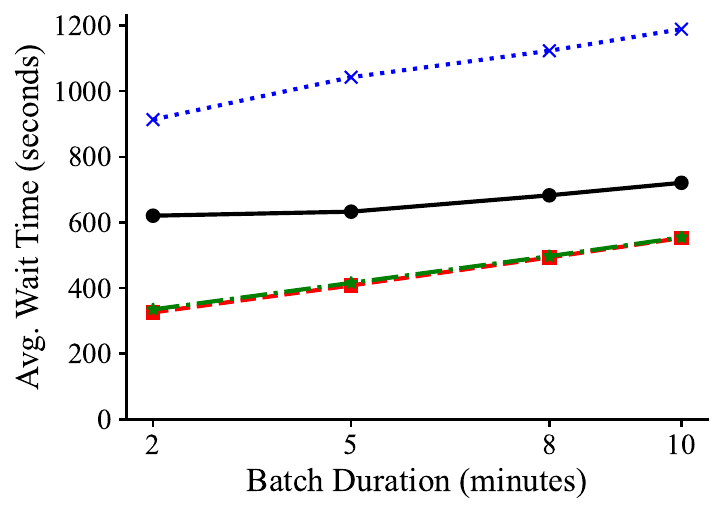} &  \includegraphics[width=0.31\linewidth]{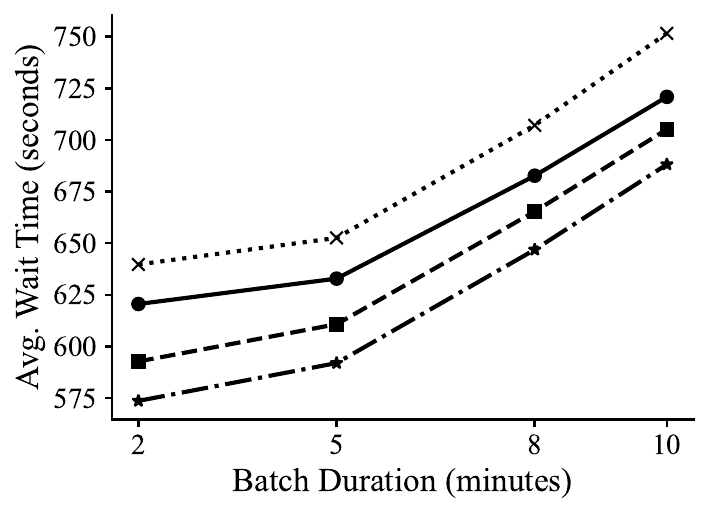} & \includegraphics[width=0.31\linewidth]{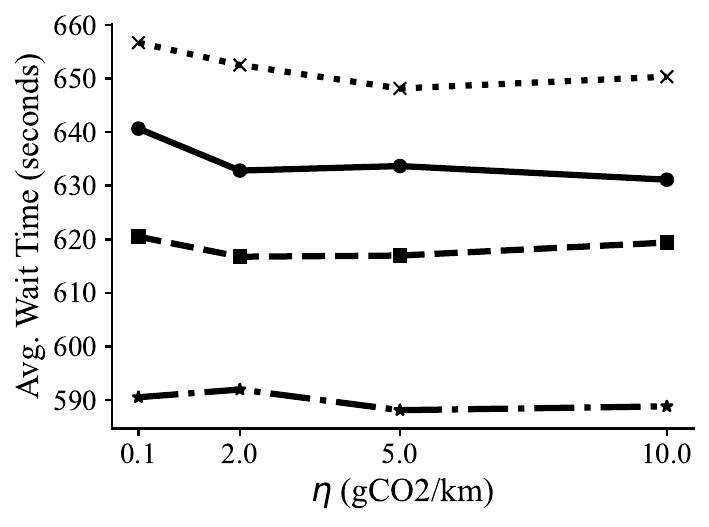}\vspace{-0.1cm}\\
  (a) \algName vs. baselines                &    (b) Impact of LEVs      &  (c) Impact of parameter $\eta$       
\end{tabular}
    \vspace{-0.2cm}
    \caption{Wait time performance: (a) wait time for different algorithms as a function of batch duration, (b) impact of the percentage of low emission vehicles (LEVs) in the fleet on \algName performance for different batch durations and $\eta$ = 5, and (c) the impact of increasing emissions for an increase in fairness, captured using $\eta$, for a batch duration of 5 minutes. Here, $\eta$ specifies extra emissions that the algorithm incurs to reduce unfairness by 1km.
    }
    \label{fig:wait-time} 
\end{figure*}

\mahsa{~\autoref{fig:wait-time} shows the impact of \algName on the wait time for riders against the baselines across various parameters. Note that \greedy targets reduce wait time and deadhead miles (or emission) by prioritizing the assignment of the closest available driver. On the other hand, \TORA is designed with explicit consideration for wait times. It incorporates mechanisms to bound the wait time experienced by the rider.}



\noindent
\textbf{Key results.}  As shown in~\autoref{fig:wait-time}~(a), \algName outperforms \LAF in terms of reducing rider wait times across all batch durations. 
Compared to \LAF, \algName shows significant performance gains, with improvements consistently exceeding~32\%.


\mahsa{~\autoref{fig:wait-time}~(b) shows that the wait time for the riders gets worse at a lower penetration of LEVs. 
For example, the increase in wait time, even at the worst point, is 4.2\% for 10\% LEVs at 10-minute batch duration.
Interestingly, increasing LEVs from 15\% to 25\% reduces the wait time, but by at most 7.5\% and at least 4.5\%.}

~\autoref{fig:wait-time}~(c) shows that increasing $\eta$ has a small effect on the rider's waiting time as well, the increase in waiting time is between 2.2\% and 3.1\% for 15\% LEVs, and the decrease in waiting time is between 6.5\% and 7.8\% for 25\% LEVs.

\noindent
\textbf{Analysis of findings.}  We next delve deeper into some results observed about \algName's impact on wait time.

\begin{itemize}[leftmargin=*]
    \item \textbf{Why does wait time for \algName increase as the batch duration increases?}  
    
    Longer batch durations delay the next assignment, causing riders to wait more, even if closer drivers become available in the meantime. This leads to an overall increase in average wait time for riders.
    

    \item \textbf{Why having more LEVs is good for wait time? How does it impact the usefulness of \algName?}  
    
    As \algName focuses on reducing emissions, it prioritizes assigning rides to LEVs, even if they are farther from the passenger, currently serving another ride request, or will only become available later. When the percentage of LEVs is low, finding nearby LEVs is challenging, leading to increased passenger wait times. However, as the proportion of LEVs increases, such as to 25\%, the algorithm has access to a larger pool of LEVs, making it easier to locate available LEVs closer to passengers, thereby reducing wait times and enhancing the algorithm's effectiveness.
     

\end{itemize}

\vspace{-0.15cm}
\begin{visionbox}{}
\noindent
\textbf{Key takeaways.} 
\emph{\algName outperforms \LAF by considering the future availability of the drivers. However, what sets \algName apart is its reduced sensitivity to batch duration and $\eta$ with respect to wait times. This allows \algName to achieve lower emissions and improved fairness.}
\end{visionbox}

\section{Conclusion}
\label{sec:conclusion}
\vspace{1mm}
In this paper, we propose \algName, a Learning-based Equity-Aware Decarbonization approach for ridesharing platforms. The primary objective of \algName is to minimize carbon emissions while ensuring a fair distribution of drivers utility. By leveraging reinforcement learning,
\algName optimizes the rider-driver matching process based on the expected future utility the driver could achieve and the projected carbon emissions for the platform without an increase in rider waiting times. 
Through extensive experiments using a publicly available real-world ridesharing dataset and a variety of problem and algorithm parameters, we demonstrate that \algName outperforms state-of-the-art baselines in terms of reduction in emissions and wait time 
while ensuring a desirable level of fairness among drivers with varying vehicle emissions.
In the future, we will focus on expanding our fairness approach to include the rider's perspective, ensuring equitable ridesharing service across different rider demographics and locations. Additionally, we plan to investigate short-term fluctuations in driver utility, complementing the long-term impacts explored in this paper, to better understand the dynamic nature of driver income and utility within ridesharing platforms.

\section{Acknowledgements}
\label{sec:acknowledgements}
This research was supported in part by NSF grants CAREER 2045641, CNS-2102963, CNS-2106299, CPS-2136199, NRT-2021693, and NGSDI-2105494.


\bibliographystyle{ACM-Reference-Format}
\bibliography{paper}

\appendix
\end{document}